# Structural, elastic, electronic, bonding, and optical properties of topological CaSn$_3$ semimetal

M. I. Naher and S. H. Naqib*

*Department of Physics, University of Rajshahi, Rajshahi 6205, Bangladesh*
*Corresponding author email: salehnaqib@yahoo.com

**Abstract**

In recent days, topological semimetals have become an active branch of materials research. The topological Weyl semimetal CaSn$_3$, belonging to the AuCu$_3$ type cubic structure, is an important electronic system to investigate both from the point of view of fundamental physics and prospective applications. In this work, we have studied the structural, elastic, mechanical, electronic, bonding, Fermi surface and optical properties of CaSn$_3$ in detail via first-principles method using the density functional theory. A comprehensive study of elastic constants and moduli shows that CaSn$_3$ possesses low level of elastic anisotropy, reasonably good machinability, mixed bonding characteristics with ionic and covalent contributions, brittle nature and relatively high Vickers hardness with a low Debye temperature. The mechanical stability conditions are fulfilled. Analysis of bond population supports the bonding nature as indicated by the elastic parameters. The bulk electronic band structure reveals clear semimetallic features with signature Dirac cone-like dispersions near the Fermi level. A pseudogap in the electronic energy density of states at the Fermi level separating the bonding and the antibonding peaks points towards significant electronic stability of cubic CaSn$_3$. The Fermi surface mostly consists of electron-like sheets with very few small hole pockets. The band structure is fairly isotropic in the *k*-space. The optical constants show interesting characteristics. The reflectivity spectra show almost non-selective behavior over a wide range of photon energy encompassing infrared to mid-ultraviolet regions. High reflectivity over wide spectral range makes CaSn$_3$ a suitable material for reflecting coating. CaSn$_3$ is an efficient absorber of ultraviolet radiation. The refractive index is very high in the infrared to visible range. All the energy dependent optical parameters exhibit clear metallic signatures and are in complete accord with the underlying bulk electronic density of states calculations.

## 1. Introduction

Topological materials, such as topological insulators and topological semimetals, have recently become a subject of great interest in physical, chemical and materials sciences [1, 2]. The basic electronic states of these compounds remain invariant under topological transformations. This implies that without changing basic topological property, one can reshape topological materials into different equivalent topological shapes. Topological materials show exciting and exotic electronic properties. Topological insulators are electrical insulators within their bulk, whereas an electric current can propagate easily on their surface [1, 2]. The surface states are induced by the topology of the bulk band structure of these quantum systems.

Topological semimetal is a newly predicted and experimentally discovered topological state of condensed matter. Though the topological system starts with the remarkable discovery of topological insulators [1, 2], in the past few decade topological semimetals have become an increasingly important area in condensed matter physics and materials science research [3, 4]. Large volume of theoretical and experimental studies on variety of topological semimetals has been done so far [5 – 10]. Topological semimetals show a rich class of exotic transport properties which are interesting from both the point of view of applications and fundamental physics encompassing materials properties to high energy particle physics.

$CaSn_3$ is a topological semimetal belonging to the $AuCu_3$ type structure exhibiting variety of interesting electronic correlations including superconductivity, heavy fermion behavior, magnetic order, and so on. Compounds with $AuCu_3$ type structure have received a great deal of attention of the scientific community [11 – 20]. Recently, through the measurement of magnetic susceptibility, electrical resistivity and specific heat, it has been revealed that $CaSn_3$ exhibits superconductivity with a superconducting transition temperature, $T_C$ = 4.2 K with typical type II behavior [21]. $CaSn_3$ also exhibits heavy fermionic character. Because of their topological electronic band structure and intrinsically low thermal conductivity, topological semimetals can be used as energy converter or thermoelectric converter, efficient catalyst, magnetic storage media and energy-efficient microelectronic components. Both quantum spin Hall and quantum anomalous Hall effects arise from topological edge states and can sustain dissipationless charge transport, which are of great potential for applications in electronics, spintronics, and quantum computation sectors. These exotic properties and potential for applications make us interested to get a deeper insight into the physical properties of $CaSn_3$.

Structural, elastic, electronic (with and without SOC), magnetic, enthalpy, vibrational, electron phonon coupling, and some thermal properties have been studied for $CaSn_3$ [22 – 24] so far. Among these, elastic properties have been explored lightly, and to the best of our knowledge, a detailed study of elastic properties including Cauchy pressure, tetragonal shear modulus, Kleinman parameter, machinability index, hardness, and anisotropy in elastic moduli are still lacking. The acoustic velocities, Debye temperature, electronic properties related to the Fermi surface, charge density distribution of atoms, Mulliken bond population analysis and optical properties of $CaSn_3$ have not yet been discussed theoretically and experimentally at all. As far as possible applications are concerned, a thorough understanding of the elastic and mechanical response of a compound is essential. This enables us to predict the behavior of the material under different conditions of pressure, stress and strain. The ductile/brittle behavior is closely associated with machinability; elastic anisotropy indices give idea about the possible mechanical failure modes. The values of elastic moduli represent the underlying bonding strength among the atoms within the crystal. The Debye temperature is linked to a number of thermo-physical properties. Information regarding frequency dependent optical constants is important in judging the potential of a material for optoelectronic applications. Therefore, in this study we undertook the $CaSn_3$ semimetal to investigate these bulk properties in details.

The rest of this paper has been designed as follows. In Section 2, we describe the computational methodology in brief. Section 3 contains the results and analysis of all the physical properties studied here. Finally, in Section 4, the theoretical results are discussed in detail and important conclusions of this study are drawn.

## 2. Computational methodology

CASTEP (Cambridge Serial Total Energy Package) code [25] has been used for the calculation of ground state energy of the material using first principles technique. This code makes use of plane wave pseudopotential method based on the density functional theory (DFT) [26, 27]. The electronic exchange-correlation energy has been calculated under the generalized gradient approximation (GGA) in the scheme due to Perdew-Burke-Ernzerhof (PBE) [28]. The interaction between the valence electrons and ion cores of Ca and Sn atoms has been represented by the Vanderbilt-type ultra-soft pseudopotential [29]. Use of ultra-soft pseudopotential saves substantial computational time with little loss of accuracy. The valence electron configurations have been taken as $3p^6 4s^2$ for Ca and $5s^2 5p^2$ for Sn atoms, respectively, in this study. In order to get the lowest energy structure, geometry optimization of $CaSn_3$ has been performed using the Broyden–Fletcher–Goldfarb–Shanno (BFGS) minimization scheme [30]. The cut off energy for the plane-wave expansion for this compound was set to be 700 eV. The Brillouin zone (BZ) integrations have been performed using the Monkhorst-Pack method [31] with a mesh size of 15×15×15 special k-points. On the other hand, the Fermi surfaces were obtained by sampling the whole BZ with the k-point mesh of size 35×35×35. Geometry optimization was performed using total energy convergence tolerance within $10^{-5}$ eV/atom, maximum lattice point displacement within $10^{-3}$ Å, maximum ionic Hellmann-Feynman force within 0.03 eV Å$^{-1}$ and maximum stress tolerance of 0.05 GPa, by using the finite basis set corrections [32]. These selected levels of tolerances produced reliable estimates of structural, elastic and electronic band structure properties with an optimum computational time.

The single crystal elastic constants, $C_{ij}$, of cubic structure were calculated based on stress-strain method [33]. A cubic crystal has three independent elastic constants ($C_{11}$, $C_{12}$ and $C_{44}$). All the elastic properties, such as the bulk modulus (B) and shear modulus (G), can be evaluated from the values of single crystal elastic constants $C_{ij}$ by using the Voigte-Reusse-Hill (VRH) approach [34, 35].

Considering matrix elements for interband optical transitions, all the optical parameters can be obtained from the complex dielectric function, $\varepsilon(\omega) = \varepsilon_1(\omega) + i\varepsilon_2(\omega)$. The imaginary part of the dielectric function, $\varepsilon_2(\omega)$, has been calculated by using the CASTEP supported formula,

$$\varepsilon_2(\omega) = \frac{2e^2\pi}{\Omega\varepsilon_0} \sum_{k,v,c} |\langle \Psi_k^c | \hat{u}.\vec{r} | \Psi_k^v \rangle|^2 \; \delta(E_k^c - E_k^v - E) \tag{1}$$

here, $\Omega$ is the unit cell volume, $\omega$ is the frequency of the incident photon, e is the charge of an electron, $\hat{u}$ is the unit vector defining the incident electric field polarization, and $\Psi_k^c$ and $\Psi_k^v$ are the conduction and valence band wave functions at a given wave-vector $k$, respectively. This formula makes use of the calculated electronic band structure. The real part of the dielectric function, $\varepsilon_1(\omega)$, has been found from the corresponding imaginary part $\varepsilon_2(\omega)$ using the Kramers-Kronig transformations. Once the values of $\varepsilon_1(\omega)$ and $\varepsilon_2(\omega)$ are known, the refractive index, the absorption coefficient, the energy loss-function, the reflectivity, and the optical conductivity can be extracted from them [36].

The Mulliken populations were investigated to understanding the bonding characteristics of CaSn$_3$ using a projection of the plane-wave states onto a linear combination of atomic orbital basis sets [37, 38]. This scheme is widely used to study charge transfers during bond formation and bond population.

3. **Results and analysis**
 *3.1 Structural Properties*

CaSn$_3$ assumes a simple cubic structure with space group $Pm\bar{3}m$ (space no. 221), where the Ca atoms occupy the simple cubic corner lattice points and Sn atoms lie on the center of each face as illustrated in Fig.1. The Ca and Sn atoms occupy the following Wyckoff positions in the unit cell [22]: Ca atoms (0,0,0) and Sn atoms (0.5, 0.5, 0). Each Ca atom is surrounded by twelve Sn atoms and each Sn atom has four nearest neighbor Ca atoms. The results of first-principles calculations of structural properties of CaSn$_3$ with their available theoretical and experimental values [21, 22] are shown in Table 1. The calculated value of lattice constant is in good agreement with previous studies.

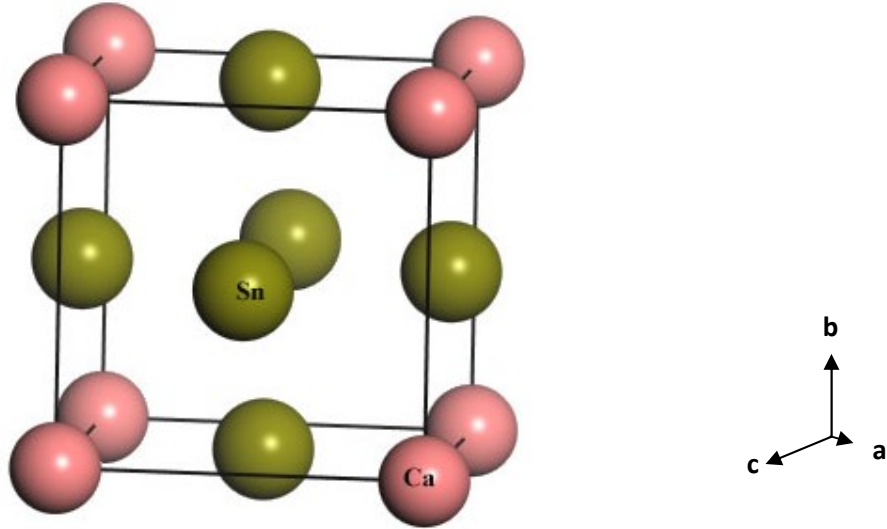

**Figure 1.** 3D Crystal structure of CaSn$_3$ unit cell.

**Table 1**
Calculated and experimental lattice constant a (Å), equilibrium volume $V_o$ (Å$^3$) and bulk modulus $B$ (GPa) of CaSn$_3$.

| Compounds | a | $V_o$ | B | Ref. |
|---|---|---|---|---|
| CaSn$_3$ | 4.77 | 108.87 | 44.44 | This |
| | 4.74 | 106.63 | - | [21]$^{Exp.}$ |
| | 4.69 | 103.43 | 50.3 | [22]$^{Theo.}$ |

*3.2 Mechanical and Elastic Properties*

Elastic properties of a solid reflect mainly the interatomic interactions and are also useful to understand various solid state properties at equilibrium state such as structural stability, bonding nature, elastic response to external stress, machinability and so on. A material with cubic symmetry has three independent elastic constants: $C_{11}$, $C_{12}$ and $C_{44}$. This is because of crystal symmetry $C_{11} = C_{22} = C_{33}$, $C_{12} = C_{23} = C_{13}$ and $C_{44} = C_{55} = C_{66}$. The calculated elastic constants are tabulated in Table 2. For mechanical stability, according to Born-Huang conditions, a cubic system requires to satisfy the following criteria [39]: $C_{11} - C_{12} > 0$; $C_{11} + 2C_{12} > 0$; $C_{44} > 0$. All the elastic constants of CaSn$_3$ satisfy the mechanical stability criteria. This indicates that CaSn$_3$ is mechanically stable.

The parameter $C' = \frac{C_{11}-C_{12}}{2}$, is a measure of crystal stiffness (the resistance to shear deformation by a shear stress applied in the (110) plane in the $[1\bar{1}0]$ direction) of a crystal. Furthermore, to calculate the Kleinman parameter ($\zeta$), isotropic bulk modulus ($B$) and shear modulus ($G$) (by the Voigt-Reuss-Hill (VRH) method), Young's modulus ($Y$), Poisson's ratio (v) and hardness ($H_V$), the following well known equations have been used [40 – 42]:

$$\zeta = \frac{C_{11} + 8C_{12}}{7C_{11} + 2C_{12}} \tag{2}$$

$$B_H = \frac{B_V + B_R}{2} \tag{3}$$

$$G_H = \frac{G_V + G_R}{2} \tag{4}$$

$$Y = \frac{9BG}{(3B + G)} \tag{5}$$

$$v = \frac{(3B - 2G)}{2(3B + G)} \tag{6}$$

$$H_V = \frac{(1 - 2v)Y}{6(1 + v)} \tag{7}$$

The resistance to shear deformation with respect to a tangential stress applied across the (100) plane in the [010] direction of a compound is parameterized by $C_{44}$. Here it is seen that $C_{44}$ is lower than $C_{11}$, which indicates that $CaSn_3$ is more easily deformed by a shear in comparison to a unidirectional compression along any of the three crystallographic directions. Larger value of $B$ compared to $G$ (Table 3) indicates that the mechanical stability will be dominated by shear modulus. The value of dimensionless Kleinman parameter generally lies in the range $0 \leq \zeta \leq 1$. According to Kleinman [43], the lower limit of $\zeta$ represents insignificant contribution of the bond bending to resist the external stress, while the upper limit corresponds to the insignificant contribution of the bond stretching/contracting to resist external applied stress. From Eqn. 2, it follows that for $C_{11} = C_{22}$, $\zeta$ reaches to its maximum value of 1. The calculated value of $\zeta$ is 0.566, from which we can predict that mechanical strength in $CaSn_3$ is mainly dominated by the bond bending contribution compared to the one that is due to bond stretching or contracting. From Table 3, we see that $B_V = B_G = B_H$. The Young's modulus, which is the ratio between tensile stress and strain, is the measurement of the resistance (stiffness) of an elastic solid to a change in its length [44, 45] and provides with a measure of thermal shock resistance. Tables 2 & 3 show that values calculated here are in very good agreement with previous theoretical estimations, except for $C_{44}$ and as a consequence the shear modulus $G$. Present calculations predicts much higher capability of $CaSn_3$ to resist shape deformation. Pugh's ratio [46 – 48] is an efficacious indicator of brittle/ductile nature of a material. Materials with $G/B$ value higher than 0.50 are brittle, otherwise it would be ductile. In case of CaSn$_3$, $G/B = 0.53$, which means that it is expected to be brittle in nature. Cauchy pressure is another indicator of brittle/ductile nature of a compound, where a compound with positive Cauchy pressure is ductile and with negative Cauchy pressure is brittle [49]. Our calculated value of Cauchy pressure of CaSn$_3$ is positive, which suggested that compound under consideration should be ductile in nature. A Poisson's ratio, $\upsilon \sim 0.31$ is another indicator of brittle and ductile threshold [50]. This implies that CaSn$_3$ is brittle in nature. We discuss these discrepancies in the next section. The value of $\upsilon = 0.25$ is the lower limit for central-force solid [51]. From the value of Poisson's ratio in Table 3, we predict that interatomic forces of CaSn$_3$ are central in nature. Poisson's ratio is also an indicator of covalent and ionic bonding in a compound. For covalent material, $\upsilon = 0.10$, whereas for ionic materials the value of $\upsilon$ is typically 0.25 [52]. The calculated Poisson's ratio of CaSn$_3$ is 0.27. This implies that ionic contribution dominates in CaSn$_3$.

The ease at which a material can be machined using cutting/shaping tools is known as machinability and it is measured by a parameter known as machinability index. Information regarding machinability of a material is becoming valuable in today's industry, because it defines the optimum economic level of machine utilization, cutting forces, temperature and plastic strain. The machinability of a material is expressed in terms of machinability index, $\mu_M$ [53]:

$$\mu_M = \frac{B}{C_{44}} \qquad (8)$$

which can also be used as a measure of plasticity [54 – 57] and lubricating nature of a material. From this equation we can say that, lower $C_{44}$ value gives better dry lubricity. A compound with large value of $B/C_{44}$ has excellent lubricating properties, lower feed forces, lower friction value, and higher plastic strain value. The $B/C_{44}$ value of CaSn$_3$ is 1.71. This implies good level of machinability. Hardness is needed to understand elastic and plastic properties of a compound. The estimated Vickers hardness, $H_V$, of CaSn$_3$ is 3.65 GPa.

**Table 2**
Calculated elastic constants, $C_{ij}$ (GPa), Cauchy pressure, $C''$ (GPa), tetragonal shear modulus, $C'$ (GPa) and Kleinman parameter ($\zeta$) for CaSn$_3$ at P = 0 GPa and T = 0 K.

| Compounds | $C_{11}$ | $C_{12}$ | $C_{44}$ | $C''$ | $C'$ | $\zeta$ | Ref. |
|---|---|---|---|---|---|---|---|
| CaSn$_3$ | 71.59 | 30.86 | 25.97 | 4.89 | 20.37 | 0.566 | This |
| | 74.00 | 32.80 | 4.60 | - | - | - | [48][Theo.] |

**Table 3**

The calculated isotropic bulk modulus $B$ (in GPa), shear modulus $G$ (in GPa), deduced by Voigt, Reuss, and Hill (VRH) approximations, and Young's modulus $Y$ (in GPa), Pugh's indicator $G/B$, Machinability index $B/C_{44}$, Poisson's ratio $\nu$ and Vickers hardness $H_V$ (in GPa) of CaSn$_3$ compound.

| Compound | B | | | G | | | Y | G/B | B/C$_{44}$ | ν | H$_V$ | Ref. |
|---|---|---|---|---|---|---|---|---|---|---|---|---|
| | B$_v$ | B$_R$ | B$_H$ | G$_V$ | G$_R$ | G$_H$ | | | | | | |
| CaSn$_3$ | 44.44 | 44.44 | 44.44 | 23.73 | 23.39 | 23.56 | 60.43 | 0.53 | 1.71 | 0.27 | 3.65 | This |
| | 46.50 | 46.50 | 46.50 | 11.00 | 6.70 | 8.80 | 24.90 | 0.19 | - | 0.41 | - | [48]$^{Theo.}$ |

It is well known that a number of physical properties such as behavior of micro-cracks in ceramics, propagation of cracks, development of plastic deformations in crystals, etc. are related to the anisotropic elastic properties of a crystal. For example, the shear anisotropic factors are the measure of degree of anisotropy in the bonding strength for atoms located in different planes. A proper elucidation of these properties has important implications in crystal physics as well as in applied engineering sciences. Anisotropic indices explain directional dependence of elastic properties of a crystal. In order to understand these properties and find mechanism to improve their durability, applications under different external environments, it is important to calculate elastic anisotropic factors of CaSn$_3$ in details.

The shear anisotropic factor for a cubic crystal can be quantified by three factors [58, 59]:
For {100} shear planes between the ⟨011⟩ and ⟨010⟩ directions, the shear anisotropic factor, $A_1$ is,

$$A_1 = \frac{4C_{44}}{C_{11} + C_{33} - 2C_{13}} \tag{8}$$

For the {010} shear plane between ⟨101⟩ and ⟨001⟩ directions the shear anisotropic factor, $A_2$ is,

$$A_2 = \frac{4C_{55}}{C_{22} + C_{33} - 2C_{23}} \tag{9}$$

For the {001} shear planes between ⟨110⟩ and ⟨010⟩ directions, the anisotropic factor, $A_3$ is,

$$A_3 = \frac{4C_{66}}{C_{11} + C_{22} - 2C_{12}} \tag{10}$$

The calculated values of these anisotropic factors are enlisted in Table 4. For an isotropic crystal $A_1 = A_2 = A_3 = 1$, while any other value is a measure of degree of anisotropy possessed by the crystal. The values of $A_1$, $A_2$ and $A_3$ show that CaSn$_3$ is moderately anisotropic and all the components of the shear anisotropic factors are equal (reflecting cubic symmetry). Zener anisotropy factor, $A$ has been calculated using following equation [41],

$$A = \frac{2C_{44}}{C_{11} - C_{12}} \tag{11}$$

Zener anisotropy factor shows same value as shear anisotropy factor of $CaSn_3$.
The universal log-Euclidean index is defined by using a log-Euclidean formula [60, 61],

$$A^L = \sqrt{\left[\ln\left(\frac{K^V}{K^R}\right)\right]^2 + 5\left[\ln\left(\frac{C_{44}^V}{C_{44}^R}\right)\right]^2} \tag{11}$$

In this scheme, the Voigt and Reuss values of $C_{44}$ is obtained from [62]

$$C_{44}^V = C_{44}^R + \frac{3}{5}\frac{(C_{11} - C_{12} - 2C_{44})^2}{3(C_{11} - C_{12}) + 4C_{44}} \tag{12}$$

and

$$C_{44}^R = \frac{5}{3}\frac{C_{44}(C_{11} - C_{12})}{3(C_{11} - C_{12}) + 4C_{44}} \tag{13}$$

Zero value of $A^L$ indicates perfect anisotropy. The values of $A^L$ range between 0 to 10.26, and for 90% of the compounds $A^L < 1$. It has been argued that $A^L$ is also an indicator regarding the layered/lamellar type of configuration [62]. Compounds with higher $A^L$ values show strong layered structural features and with lower $A^L$ values show non layered structure. From the comparatively lower value of $A^L$ we can predict that our compound exhibits moderately layered type of configuration. We have also calculated the universal anisotropy factor, $A^U$. The universal anisotropy index $A^U$, equivalent Zener anisotropy measure, $A^{eq}$, anisotropy in compressibility $A_B$ and anisotropy in shear $A_G$ (or $A^C$) for the crystal are calculated using following standard equations [60, 63 – 64]:

$$A^U = 5\frac{G_V}{G_R} + \frac{B_V}{B_R} - 6 \geq 0 \tag{14}$$

$$A^{eq} = \left(1 + \frac{5}{12}A^U\right) + \sqrt{\left(1 + \frac{5}{12}A^U\right)^2 - 1} \tag{15}$$

$$A^C = \frac{G^V - G^R}{2G^H} \tag{16}$$

$$A_B = \frac{B_V - B_R}{B_V + B_R} \tag{17}$$

Because of the simplicity in comparing the plurality of anisotropy factors definition for specific planes in crystals, a singular anisotropy index has become attractive. Ranganathan and Ostoja-Starzewski [63] defined the universal anisotropy index, which provides a singular measure of anisotropy. $A^U$ is called *universal* because of its applicability to all sorts of crystal symmetries. The condition for an isotropic crystal is $A^U = 0$. Derivation from this value, which must be positive, suggests presence of anisotropy. $A^U$ for CuSn$_3$ is 0.073, which slightly derivates from zero. That means CuSn$_3$ possess small amount of anisotropy in elastic properties. Equivalent Zener anisotropy measure is calculated using Eqn. 15. For isotropic crystal, $A^{eq} = 1$. The calculated value of $A^{eq}$ is 1.279, predicting that CaSn$_3$ is moderately anisotropic. However, Chung and Buessem [65] refrained from extending $A^C$ to crystals with lower symmetries because, in addition to the shear modulus, the bulk modulus influences the anisotropy of crystals having other than cubic symmetry [65]. It is notable that compared to other measures, $A^C$ predicts much lower value (0.007) of anisotropy index. Since $A_B$ is the percentage of the measurement of anisotropy in compressibility, a zero value of $A_B$ for CaSn$_3$ with cubic structure proves that bulk modulus has no influences to the anisotropic elastic and mechanical properties.

The bulk modulus along *a*, *b* and *c* axis and anisotropies of the bulk modulus can be defined as [59]:

$$\begin{aligned} B_a &= a\frac{dP}{da} = \frac{\Lambda}{1 + \alpha + \beta} \\ B_b &= a\frac{dP}{db} = \frac{B_a}{\alpha} \\ B_c &= c\frac{dP}{dc} = \frac{B_a}{\beta} \end{aligned} \tag{18}$$

$$\begin{aligned} A_{B_a} &= \frac{B_a}{B_b} = \alpha \\ A_{B_c} &= \frac{B_c}{B_b} = \frac{\alpha}{\beta} \end{aligned} \tag{19}$$

where, $\Lambda = C_{11} + 2C_{12}\alpha + C_{22}\alpha^2 + 2C_{13}\beta + C_{33}\beta^2 + 2C_{33}\alpha\beta$ and for cubic crystals $\alpha = \beta = 1$. $A_{B_a}$ and $A_{B_c}$ represent anisotropies of bulk modulus along the $a$ axis and $c$ axis with respect to $b$ axis, respectively. A value of $A_{B_a} = A_{B_c} = 1$ indicates isotropy. This means that bulk modulus is isotropic in CaSn$_3$. Calculated bulk moduli of CaSn$_3$ along different crystallographic axes are equal; 133.31 GPa. This value is different from the isotropic bulk modulus and is larger.

**Table 4**
Shear anisotropic factors ($A_1$, $A_2$ and $A_3$), Zener anisotropy factor $A$, universal log-Euclidean index $A^L$, the universal anisotropy index $A^U$, equivalent Zener anisotropy measure $A^{eq}$, anisotropy in shear $A_G$ (or $A^C$), anisotropy in compressibility $A_B$, anisotropy in bulk modulus, and anisotropic bulk modulus (in GPa) for CaSn$_3$.

| Compound | $A_1$ | $A_2$ | $A_3$ | $A$ | $A^L$ | $A^U$ | $A^{eq}$ | $A_G$ | $A_B$ | $A_{B_a}$ | $A_{B_c}$ | $B_a$ | Layered | Ref. |
|---|---|---|---|---|---|---|---|---|---|---|---|---|---|---|
| CaSn$_3$ | 1.28 | 1.28 | 1.28 | 1.28 | 0.154 | 0.073 | 1.279 | 0.007 | 0 | 1 | 1 | 133.31 | No | This |

### 3.3 Debye Temperature and Anisotropy of Acoustic Velocities

Debye temperature is a fundamental lattice dynamical and thermal parameter. A large number of thermo-physical properties of a solid like, thermal conductivity, lattice vibration, interatomic bonding, melting temperature, coefficient of thermal expansion and specific heat can be estimated from Debye temperature. The vacancy formation energy, electron-phonon coupling constant and superconducting transition temperature of superconductors are related to Debye temperature. A compound with larger Debye temperature has a higher melting temperature, greater hardness and stronger interatomic bonding strength.

Using the estimated elastic moduli, sound velocities can be determined from following equations [66 – 68]. The average sound velocity $v_a$ can be determined from the transverse and longitudinal ($v_t$ and $v_l$, respectively) sound velocities as follows,

$$v_a = \left[\frac{1}{3}\left(\frac{2}{v_t^3} + \frac{1}{v_l^3}\right)\right]^{-\frac{1}{3}} \quad (20)$$

The transverse velocity $v_t$ is obtained from,

$$v_t = \sqrt{\frac{G}{\rho}} \tag{21}$$

where, $\rho$ is the density of the material. The longitudinal wave velocity, $v_l$, is calculated from,

$$v_l = \sqrt{\frac{B + 4G/3}{\rho}} \tag{22}$$

The Debye temperature, $\Theta_D$, is evaluated from [69]:

$$\Theta_D = \frac{h}{k_B}\left(\frac{3n}{4\pi V_0}\right)^{1/3} v_a \tag{23}$$

where, $h$ is Planck's constant, $k_B$ is the Boltzmann's constant, $V_0$ is the volume of unit cell and $n$ defines the number of atoms within the unit cell.

The calculated Debye temperature $\Theta_D$ for CaSn$_3$ along with the sound velocities $v_l$, $v_t$, and $v_a$ are listed in Table 5 together with the experimental values where available [21].

To understand the anisotropic nature of sound velocities, they should be measured in different propagation directions. Since, in an anisotropic solid there are only certain crystallographic directions along which elastic waves can propagate in pure longitudinal and transverse modes. For cubic symmetry, the pure transverse and longitudinal modes can be found for [001], [110] and [111] directions; in other directions, the sound propagating modes are the quasi-transverse or quasi-longitudinal. For cubic crystal the acoustic velocities in the principle directions can be expressed as [70]:

$$[100]v_l = \sqrt{C_{11}/\rho} \; ; \; [010]v_{t1} = [001]v_{t2} = \sqrt{C_{44}/\rho} \; ; \; [110]v_l = \sqrt{(C_{11} + C_{12} + 2C_{44})/2\rho} \; ; \tag{24}$$

$$[1\bar{1}0]v_{t1} = \sqrt{(C_{11} - C_{12})/\rho} \; ; \; [001]v_{t2} = \sqrt{C_{44}/\rho} \; ; \; [111]v_l = \sqrt{(C_{11} + 2C_{12} + 4C_{44})/3\rho} \; ;$$

$$[11\bar{2}]v_{t1} = v_{t2} = \sqrt{(C_{11} - C_{12} + C_{44})/3\rho}$$

where $v_{t1}$ and $v_{t2}$ are the first transverse mode and the second transverse mode, respectively.

From these equations we can say that a compound with the value of small density $\rho$ and large elastic constants has large sound velocities. The calculated sound velocities along different directions of CaSn$_3$ are shown in Table 6. The different values of sound velocities in different directions indicate lattice dynamical anisotropy in CaSn$_3$. For example, the longitudinal sound velocities along [100], [010] and [001] directions are determined from C$_{11}$, and C$_{44}$ corresponds to the transverse modes [71].

**Table 5**

Density $\rho$ (in g/cm$^3$), transverse velocity $v_t$ (in ms$^{-1}$), longitudinal velocity $v_l$ (in ms$^{-1}$), average elastic wave velocity $v_a$ (in ms$^{-1}$) and Debye temperature $\Theta_D$(K), for CaSn$_3$.

| Compounds | $\rho$ | $v_t$ | $v_l$ | $v_a$ | $\Theta_D$ | Ref. |
|---|---|---|---|---|---|---|
| CaSn$_3$ | 6.04 | 1974.47 | 3542.83 | 2196.83 | 212.85 | This |
| | - | - | - | - | 175 &211 | [21]$^{Exp.}$ |

**Table 6**

Anisotropic sound velocities (in ms$^{-1}$) of CaSn$_3$ along different crystallographic directions.

| Anisotropic sound velocities | | CaSn$_3$ |
|---|---|---|
| [111] | $[111]v_l$ | 3617.02 |
| | $[11\bar{2}]v_{t1,2}$ | 1918.07 |
| [110] | $[110]v_l$ | 3574.02 |
| | $[1\bar{1}0]v_{t1}$ | 2596.09 |
| | $[001]v_{t2}$ | 2072.99 |
| [100] | $[100]v_l$ | 3441.83 |
| | $[010]v_{t1}$ | 2072.99 |
| | $[001]v_{t2}$ | 2072.99 |

### *3.4 Electronic Density of States and Band Structure*

The electronic band structure, as a function of energy ($E$-$E_F$), along different high symmetry directions ($\Gamma$-*X*-*M*-*X*-*R*-$\Gamma$) in the first BZ has been calculated at zero pressure and temperature. Figure 2(a) shows the band structure of CaSn$_3$. The horizontal broken line indicates the Fermi level. The total number of bands is 57. From Fig. 2, we can see that there is no band gap at the Fermi level, due to the overlap of valence band and conduction band. This indicates metallic nature of CaSn$_3$. The bands crossing the Fermi level are shown in different colors with their corresponding band numbers. There are two bands that cross the Fermi level (bands 26 and 27). Band 27 reveals both electron-like and hole-like features along different directions in the Brillouin zone.

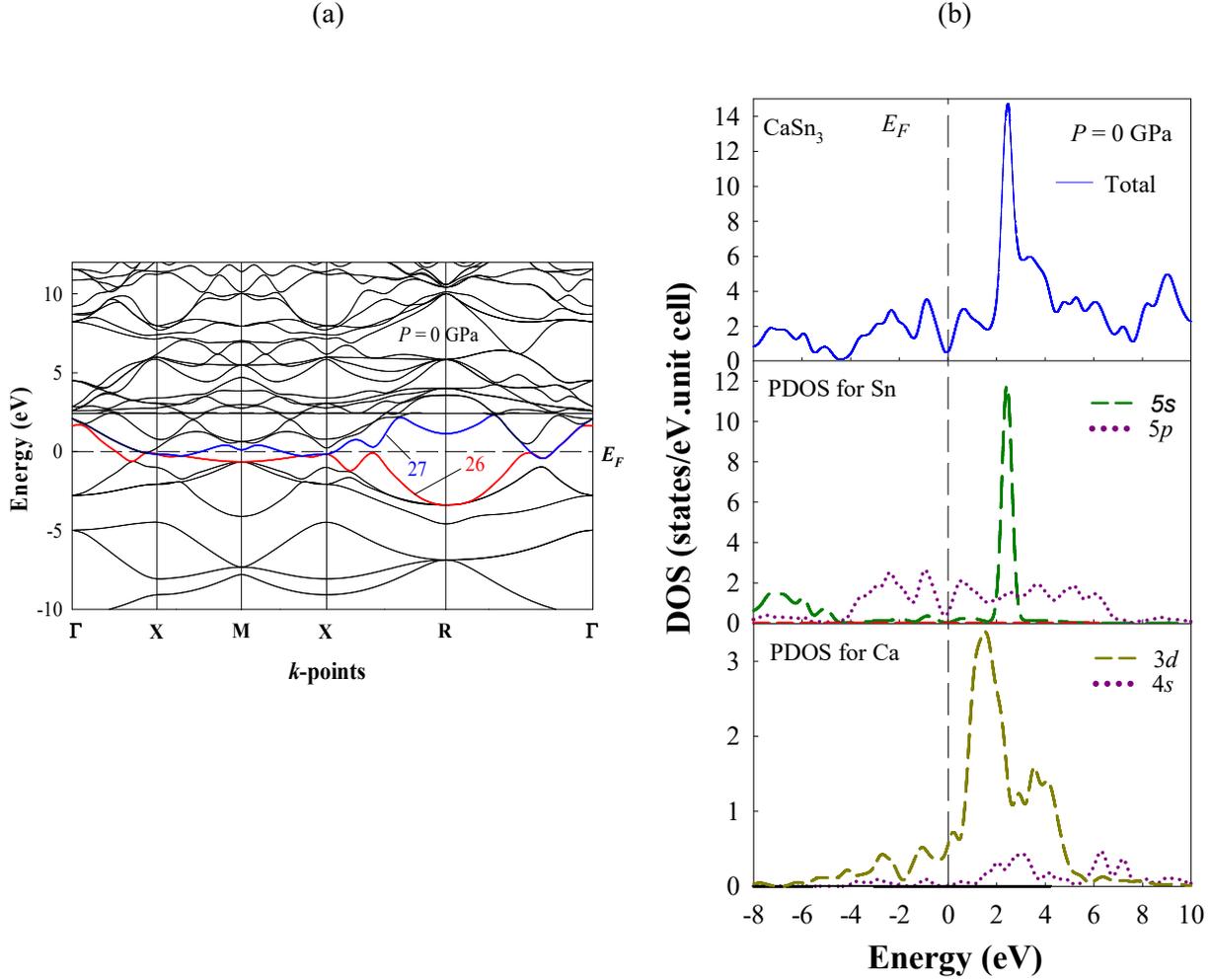

**Figure 2.** (a) Electronic band structure and (b) total and partial electronic density of states (PDOSs and TDOSs) of $CaSn_3$ along several high symmetry directions of the Brillouin zone at P = 0 GPa.

The calculated total and partial density of states (TDOS and PDOS, respectively) of $CaSn_3$ at zero pressure and temperature is shown in Fig. 2(b). The Fermi level, $E_F$ is denoted by vertical broken line. In order to understand the contribution of each atom in the TDOS, we have also calculated partial DOS of both Ca and Sn atoms in $CaSn_3$. The non-zero values of TDOS at the Fermi level indicate that $CaSn_3$ will exhibit metallic electrical conductivity. TDOS value of $CaSn_3$ at the Fermi level is 0.57 states per eV per unit cell. Near the Fermi level, the main contribution comes from $3d$ and $5p$ state of Ca and Sn atom, respectively. There is significant hybridization between these electronic sates. Thus the electronic and bonding nature of $CaSn_3$ is mainly dominated by these orbitals. It is observed that the $4s$ orbital of Ca has negligible contribution to the TDOS at the Fermi level. Electronic stability of a compound depends on the location of Fermi level and on the value of $N(E_F)$ [72, 73]. Compounds with lower values of $N(E_F)$ are more stable. The electronic stability is also related to the presence of a pseudogap or quasi-gap in the TDOSs in the vicinity of the Fermi level [74, 75]. This gape is the separation of bonding and nonbonding/antibonding states. For CaSn$_3$, Fermi level lie almost on the pseudogap. A sharp and relatively narrow band due to the 5$s$ electronic states of Sn atom contribute significantly at energies 2-3 eV in the conduction band. We predict that this high DOS peak should play notable role in determining optical behavior of the crystal. Broad electronic states due to the 5$p$ orbitals of Sn also contribute in the formation of the conduction band over an extended energy range 0-6 eV. In comparison to Sn orbitals, the role played by 3$d$ and 4$s$ electrons of Ca is minor in forming the conduction and valence bands of CaSn$_3$. It is interesting to note that the electronic dispersion of band 26 near the $X$ point (along $\Gamma$ - $X$ and $X$ – $R$) show linear characteristics. These are non-trivial electronic structures demonstrating the line nodes in the absence of spin-orbit coupling and are signatures of the band structure of topological electronic state.

### *3.5 Fermi Surface of CaSn$_3$*

In condensed matter physics, the Fermi surface is important to understand the behavior of occupied and unoccupied electronic states of a metallic material at low temperatures. Electronic, optical, thermal and magnetic properties of a material are strongly dependent on Fermi surface topology. 3D plot of the Fermi surfaces of CaSn$_3$ is shown in Fig. 3. Fermi surfaces are constructed form band number 26 and 27 of CaSn$_3$. It is seen that CaSn$_3$ contains both electron- and hole-like sheets. For band 26, there is an electron-like sheet around the $\Gamma$-point. For band 27, electron-like sheets appear along the $\Gamma$ – $R$ path. Hole-like sheets appear around the $X$-point.

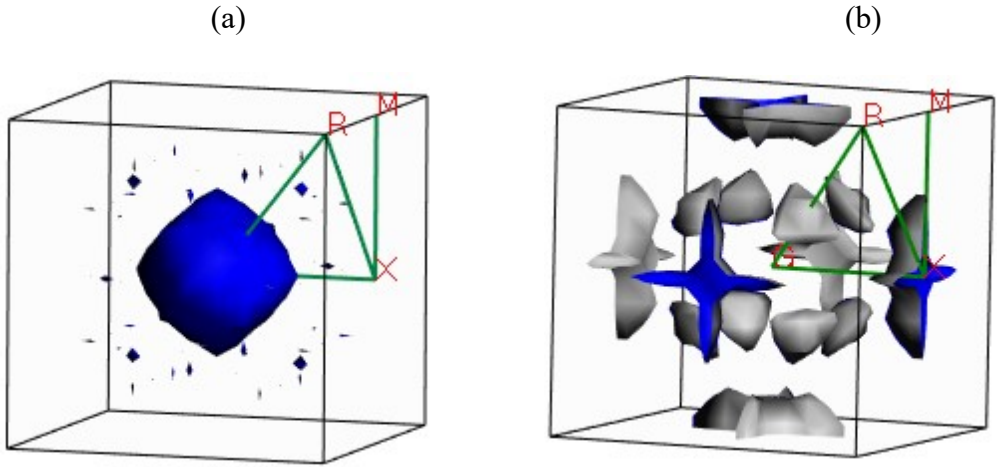

(a)        (b)

**Figure 3.** Fermi surface for bands (a) 26 and (b) 27of CaSn$_3$, respectively.

*3.6 Electronic Charge Density*

In order to understand total charge density distribution in CaSn$_3$, we have studied electronic charge density of CaSn$_3$ along different crystallographic planes. A colored scale on the right hand side of Fig. 4 shows total electron density. High charge (electron) density is indicated by blue color and low charge (electron) density is indicated by red color. From charge density map, we can see that Sn atoms have high electron density compared to Ca atoms in CaSn$_3$. Therefore, ionic bonding is expected between Ca and Sn atoms. This agrees with the result we get from the Poisson's ratio calculation.

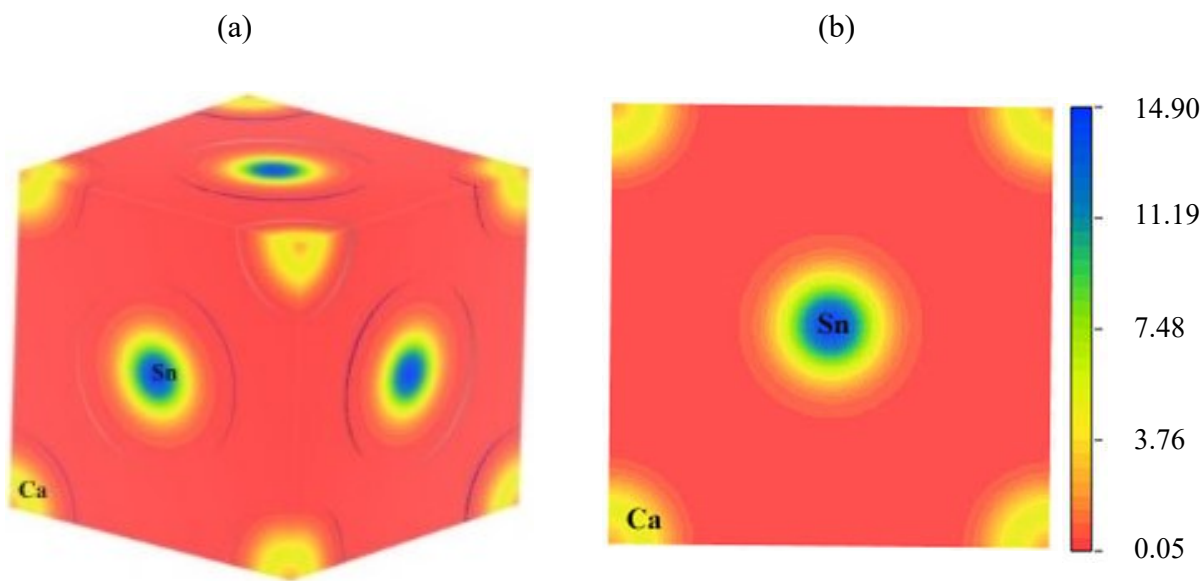

**Figure 4.** The electronic charge density map for CaSn$_3$ in (a) 3D and (b) in the (110) plane.

*3.7 Bond population analysis*

To explore the bonding nature (ionic, covalent or metallic) and effective valence of an atom in the molecule in greater depth, the Mulliken bond populations [76] are studied. The results of this analysis for CaSn$_3$ are summarized in Table 6. The amount of missing valence charges in a projection is represented by the Charge spilling parameter. The lower value of charge spilling parameter, which is 0.19%, indicates a good representation of electronic bonds. Total charge of Sn atom is much larger than Ca atom, which mainly comes from 4$d$ orbitals. Atomic charges for Ca and Sn atoms in CaSn$_3$ are 0.88 and -0.29 electron, respectively. Both are deviated from purely ionic state of Ca and Sn, which are +2 and +4, respectively. This reflects the presence of some covalent bonding character in CaSn$_3$. For CaSn$_3$, electrons are transferred from Ca to Sn atoms. From which one can predict that ionic bond dominates between Ca and Sn in CaSn$_3$. From orbital charge value we can say that these electrons mainly come from 5$p$ orbital of Ca.

Effective valence is the difference between formal ionic charge and the Mulliken charge on the anion species in the crystal [38].To understand the degree of covalency and/or iconicity, the effective valence of $CaSn_3$ is calculated. The zero value of effective valence indicates perfect ionic bond and values greater that zero indicate increasing level of covalency. Effective valence for Ca in $CaSn_3$ is +1.12. This implies that both ionic and covalent bonds are present in $CaSn_3$.

It is recognized that the Mulliken bond population analysis sometimes gives results in contradiction to chemical intuition, because of its strong basis set dependency. On the other hand, Hirshfeld population analysis (HPA), with practically no basis set dependence, provides us with more meaningful result. Keeping this point in mind, we have determined Hirshfeld charge of $CaSn_3$ using the HPA. HPA shows somewhat opposite result compared to the Mulliken charge. That has been also found for other cases [77]. Hirshfeld analysis shows atomic charge of Ca and Sn is -0.03 and +0.01 electronic charge, respectively. Hirshfeld charge predicts that electrons are transferred from Sn to Ca atoms. This is opposite result that we get from Mulliken charge analysis. At the same time, it should be noted that both the approaches predict covalent bonding between Sn and Ca. We have also calculated effective valence of $CaSn_3$ using Hirshfeld charge, which is +3.99 electron. This value is much larger than the value we get from Mulliken charge analysis. Thus we can say that compared to MPA, HPA predicts a higher degree of covalency in $CaSn_3$.

**Table 6**
Charge spilling parameter (%), orbital charges (electron), atomic Mulliken charges (electron), effective valence (electron) and Hirshfeld charge (electron) in $CaSn_3$.

| Compounds | Species | Charge spilling | s | p | d | Total | Mulliken charge | Effective valence | Hirshfeld charge | Effective valence |
|---|---|---|---|---|---|---|---|---|---|---|
| CaSn₃ | Ca | 0.19 | 2.21 | 6.00 | 0.91 | 9.12 | 0.88 | +1.12 | -0.03 | |
| | Sn | | 1.77 | 2.52 | 10.00 | 14.29 | -0.29 | | 0.01 | +3.99 |

*3.8 Optical Properties*

Optical properties are important to understand the response of a material to incident electromagnetic wave. They are also important to explore possible optoelectronic and photovoltaic device applications. In this regard the response of a compound to infrared, visible and ultraviolet spectra is particularly pertinent. Various energy dependent (frequency) optical parameters, namely real and imaginary part of dielectric constants, $\varepsilon_1(\omega)$ and $\varepsilon_2(\omega)$, respectively, real part of refractive index $n(\omega)$, extinction coefficient $k(\omega)$, loss function $L(\omega)$, real and imaginary parts of the optical conductivity ($\sigma_1(\omega)$ and $\sigma_2(\omega)$, respectively), reflectivity $R(\omega)$, and the absorption coefficient $\alpha(\omega)$, are calculated to explore the response of CaSn$_3$ to incident photons. The calculated optical constants of CaSn$_3$ for photon energies up to 20 eV with electric field polarization vectors along [100] direction is shown in Fig. 5. The electronic band structure properties of CaSn$_3$ show that the compound is metallic in nature. Therefore, the Drude damping correction is required [78, 79]. We have used a screened plasma energy of 10 eV and a Drude damping of 5 eV to initiate the calculation of the optical constants of CaSn$_3$.

Optical parameters of a material can be derived from the complex dielectric function, $\varepsilon(\omega)$. Generally, intraband and interband optical transitions determine $\varepsilon(\omega)$. In this calculation the indirect interband transitions are ignored since it involves phonons and has a small scattering cross section [80] compared to that for direct transition for which phonon scattering is not required to conserve momentum. Fig. 5(a) shows the real and imaginary part of dielectric constant. The static dielectric constant $\varepsilon_1(0)$ is an important optical parameter. It is inversely proportional to band gap value. From Fig. 5(a) we see that the compound under study is metallic, which is also evident from variety of other analyses. Metallic reflection characteristics are exhibited in the range $\varepsilon_1 < 0$.

The refractive index is a complex parameter, $N(\omega) = n(\omega) + ik(\omega)$; where $k(\omega)$ is known as extinction coefficient. Real $n(\omega)$ and imaginary $k(\omega)$ part of refractive index of CaSn$_3$ is shown in Fig. 5(b). The maximum refractive index for CaSn$_3$ is obtained at zero energy and decreases with increasing energy. The peak in the refractive index is established at the transition of an electron from valence band to the conduction band. A compound with larger values of extinction coefficient indicates that the compound has high ability to absorb light. It is instructive to note that CaSn$_3$ has very high value of $n(\omega)$ at low energies covering infrared to visible regions. This has important practical consequence. High refractive index materials have desired optical characteristics for light emitting and optoelectronic display devices.

Photoconductivity spectrum is shown in Fig. 5(c). Photoconductivity starts with zero photon energy, which indicates that the material has no band gap as evident from band structure and TDOS calculations (Fig. 2). The photoconductivity and hence the electrical conductivity of the material increases as a result of increasing energy. It reaches to its maximum value at 1.37 eV photon energy, decreases gradually with further increase in energy, and tends to zero at around 20 eV.

Reflectivity is an important parameter to understand the applicability of a system as a coating material for different optical regions. The reflectivity spectrum of CaSn$_3$ is shown in Fig. 5(d). Reflectivity values of CaSn$_3$ are shown in the range 0 to 20.6 eV photon energy. The reflectivity spectrum is interesting in certain aspects. For a very broad spectral range from infrared to mid-ultraviolet, the average value of $R(\omega)$ is ~ 60%, which does not fall below 50% in this region. $R(\omega)$ shows almost nonselective behavior over this broad energy range.

The absorption coefficient is an important parameter to understand the optimum solar energy conversion efficiency of a material. It also reveals the electrical nature of the material, whether it is metallic, semiconducting or insulating. Fig. 5(e) depicts the absorption coefficient spectra of $CaSn_3$. As it is seen from this figure that optical absorption sets in at zero photon energy, therefore $CaSn_3$ has no optical band gap. The onset of optical absorption is due to the free-electrons within the conduction band. The absorption coefficient is quite high in the spectral region from ~ 3 to 10 eV, peaking around 5.29 eV. It decreases sharply at ~ 14.5 eV in agreement with the position of the loss peak.

Fig. 5(f) shows frequency dependent energy loss function. Loss function of a material is also related to absorption and reflection characteristics of a material. Study of energy loss function $L(\omega)$, is useful for understanding the screened charge excitation spectra, especially the collective excitations produced by a swift electron traversing a solid. The loss spectrum also arises due to photon absorption with appropriate energy which can give rise to excitations of collective charge oscillations called the plasmons. The peak in $L(\omega)$ spectra appears at a particular incident light frequency (energy), known as the bulk screened plasma frequency. It is observed that the peak of $L(\omega)$ is located at 14.70 eV. Sharpe peak represents the abrupt reduction in reflectivity and absorption of $CaSn_3$ (see Figs. 5 (d) &(e)). If the incident light energy is greater than this, $CaSn_3$ will be transparent and will change its response from metallic to dielectric. The energy at which $L(\omega)$ is maximum, is known as plasmon energy and this particular value corresponds to the zero crossing point of the real part of the dielectric constant, $\varepsilon_1(\omega)$ (Fig. 5a).

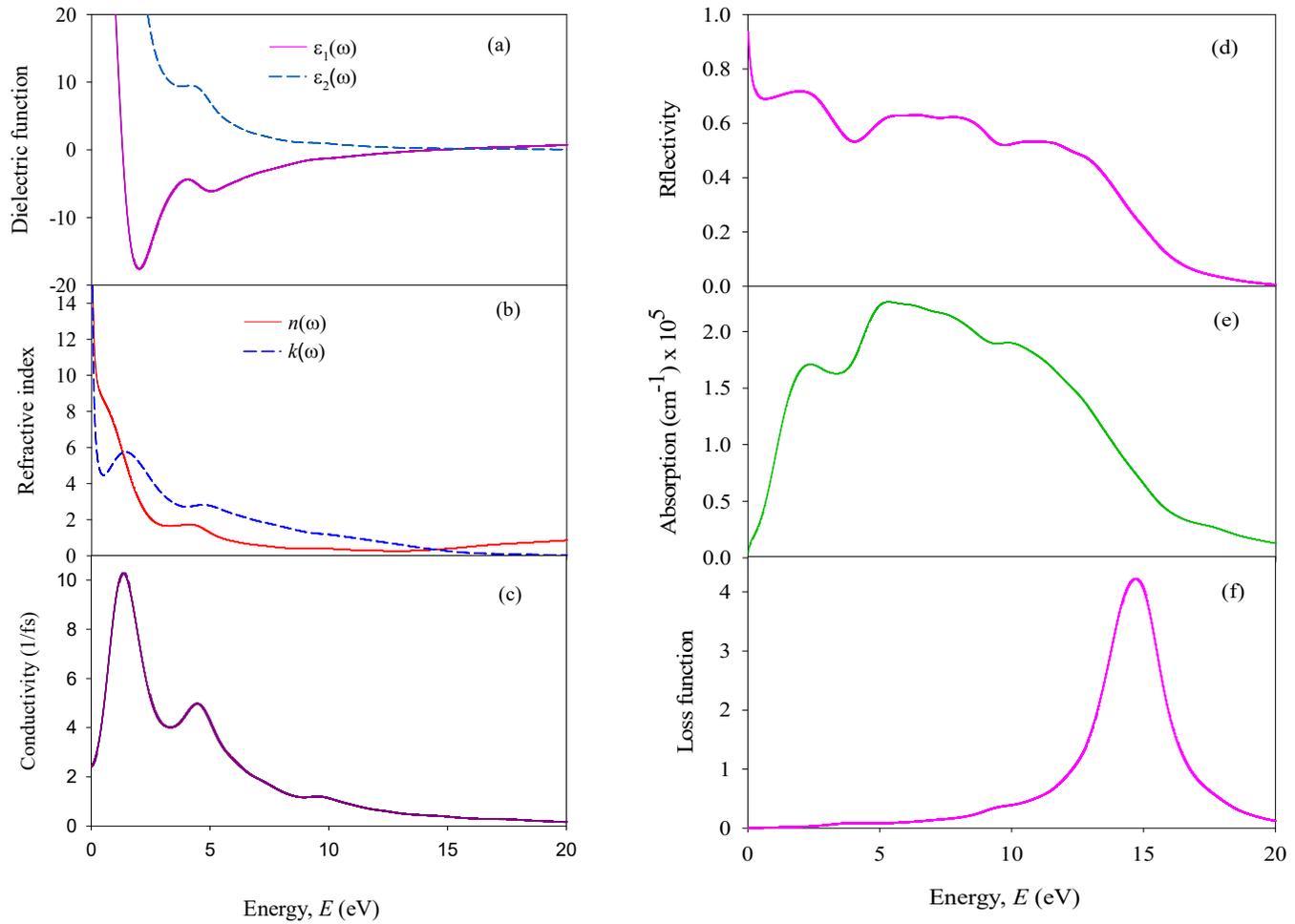

**Figure 5.** (a) $\varepsilon_1(\omega)$ and $\varepsilon_2(\omega)$, (b) $n(\omega)$ and $k(\omega)$, (c) $\sigma(\omega)$, (d) $R(\omega)$, (e) $\alpha(\omega)$ and (f) $L(\omega)$ of $CaSn_3$.

## 4. Discussion and conclusions

Detailed analysis of structural and elastic properties of $CaSn_3$ in the cubic symmetry reveals that this topological semimetal is fairly isotropic in nature, both structurally and electronically. The optimized geometry agrees well with previous experimental and theoretical results [21, 22]. The magnitude of the various elastic constants and moduli indicates that the mechanical failure mode of $CaSn_3$ is dominated by the shear deformation. The positive Cauchy pressure implies that $CaSn_3$ is expected to be ductile. Poisson's ratio of $CaSn_3$, on the other hand, suggests that this compound is brittle in nature. The analysis of charge density distribution and bond population analyses indicate that both covalent and ionic bondings are present in $CaSn_3$. This is further supported by the PDOS features, where significant hybridization between Sn5$p$ and Ca 3$d$ electronic orbitals are found. This indicates towards a tendency to the formation of covalent bonding between these two atomic species. Covalent bonding with directional character makes a material brittle. The positive Cauchy pressure for $CaSn_3$ can be misleading since corrections due to many-body interaction between atoms and electron gas are not taken into account in details in determining the elastic constants [81]. The values of the estimated single crystal elastic constants, $C_{11}$ and $C_{12}$ show excellent agreement with previous theoretical study [48]. But the estimated values of $C_{44}$ show large deviation. We note that the very small value of $C_{44}$ (4.60 GPa) predicted in Ref. 48 would result in an unrealistically high value of the machinability index and raises question regarding its accuracy. Low value of the universal log-Euclidean index ($A^L$) implies that the bonding strengths in $CaSn_3$ are fairly isotropic along different directions within the crystal and the compound does not show layered characteristics. It is instructive to note that high degree of machinability and significant Vickers hardness of fairly isotropic $CaSn_3$ are comparable with variety of layered binary and ternary compounds including the widely studied MAX phase nanolaminates [77, 82 – 86].

The Debye temperature, one of the most important thermo-physical parameters in solids, of $CaSn_3$ was determined using the calculated single crystal elastic constants. In variety of systems with diverse class of dominant chemical bondings and electronic band structure features, this method has proven to result in reliable values of the $\Theta_D$ [66, 84, 86 – 90]. Previously estimated values of $\Theta_D$ from the fits to the resistivity and heat capacity data yielded almost same value of the Debye temperature [21] (Table 5) as the value obtained here. The low estimated value of $\Theta_D$ implies that lattice thermal conductivity of $CaSn_3$ is expected to be low as well. This, together with a low EDOS at the Fermi level, predicts that $CaSn_3$ will exhibit relatively low overall thermal conductivity at all temperatures.

The gross feature of the electronic band structure and electronic energy density of states found in this study agree well with previous results [21]. The electronic density of states at the Fermi level ($N(E_F) \sim 0.80$ states/eV) shows total agreement with the previously calculated value of 0.82 states/eV [21]. Two particular bands (marked 26 and 27 in Fig. 2a) cross the Fermi level demonstrating the metallic character of $CaSn_3$. These two bands exhibit additional interesting features - certain parts of these bands are quasilinear in nature. Such dispersions are signature characteristics of topological systems. Dirac cone-like dispersions are seen in band 26 along the $\Gamma$ - $X$ direction close to the $X$ point, and in band 27 along the $R$ - $\Gamma$ direction close to the $\Gamma$ point near the Fermi energy. The Fermi surface of $CaSn_3$ is derived from these two particular bands and different Fermi sheets along different directions in the $k$-space show primarily electronic character including a few small hole pockets.

The optical constants spectra show close agreement with the underlying electronic band structure. Both absorption coefficient and optical conductivity show metallic character. $CaSn_3$ strongly absorbs ultraviolet radiation. The compound under study also exhibits almost non-selective reflectivity over a wide range of photon energy. This material has high reflectivity for spectral regions covering the infrared to mid-ultraviolet radiations and can be used as an efficient solar reflector. Furthermore, materials with high reflectivity in the infrared energies can be utilized to confine thermal energy and to minimize heat loss to the environment. The absorption coefficient of $CaSn_3$ is quite high in the ultraviolet region. Materials having high absorbance in the ultraviolet can be employed as coating to protect systems which are prone to ultraviolet radiation induced photo disintegration or photo-oxidation. All these features hold promise for potential applications. Moreover, the refractive index of $CaSn_3$ is very high in the visible region. The high refractive index of $CaSn_3$ in the infrared and visible region shows that its application in the optimization of various display systems, such as QLED, LCD and OLED, might be possible.

As mentioned in the preceding section (Section 1) $CaSn_3$ is a superconductor with a transition temperature of 4.2 K. This investigation reveals several features which can be highly relevant to superconductivity. For example, the low value of Debye temperature suggests that the lattice is quite compressible. For such a solid, application of pressure would result in a significant increase in $\varTheta_D$ which should enhance the superconducting $T_c$ for a superconductor where Cooper pairing is mediated by phonons. More importantly, application of pressure would result in a shift in the position of the Fermi level. This, for $CaSn_3$, will increase the $N(E_F)$ significantly since at ambient condition $N(E_F)$ is located at a deep valley in the EDOS. Any energy shift, positive or negative, would lead to an increase in the $N(E_F)$ (Fig. 2b). The electron-phonon coupling constant varies linearly with $N(E_F)$ for a given electron-phonon interaction energy. Therefore, we predict that pressure would increase $T_c$ significantly of $CaSn_3$. The same effect can be induced via chemical means with doping with suitable dopants.

It is perhaps worth stating that the surface electronic states are one of the defining features of topological electronic systems which arise from spin orbit coupling (SOC). In this paper we have concentrated solely on the bulk electronic, optical and mechanical properties of cubic $CaSn_3$. Variety of earlier studies on diverse class of materials have clearly showed that [77, 88, 90 – 93], as far as optimization of the cell structure, bulk elastic constants, bonding and bulk optical properties are concerned, inclusion of SOC only has a minimal bearing. Within the bulk electronic band structure, SOC mainly manifests itself in certain split bands with splitting energy of the order of tens of meV. Therefore, we have not incorporated SOC for our bulk elastic, electronic and optoelectronic calculations.

To summarize, we have investigated the elastic, mechanical, bonding, bulk electronic, Fermi surface and optoelectronic properties of cubic topological semimetal $CaSn_3$. The elastic, bonding, and optoelectronic properties are studied in details for the first time. $CaSn_3$ possesses several attractive mechanical and optoelectronic features which are suitable for engineering and device applications. We hope that this study will inspire researchers to investigate this interesting material in further details in future, both theoretically and experimentally.

**Acknowledgements**

S. H. N. acknowledges the research grant from the Faculty of Science, University of Rajshahi, Bangladesh, which partly supported this work.

**Data availability**

The data sets generated and/or analyzed in this study are available from the corresponding author on reasonable request.

**Author Contributions**

S. H. N. designed the project and wrote the manuscript. M. I. N. performed the theoretical analysis. Both the authors reviewed the manuscript.

**Additional Information**

**Competing Interests**

The authors declare no competing interests.